\documentclass{aip-cp}

\usepackage[numbers]{natbib}
\usepackage{rotating}
\usepackage{graphicx}

\begin{document}

\title{Unidentified Fermi LAT Transients Near The Galactic Plane}

\author{Dirk Pandel}
\affil{Department of Physics, Grand Valley State University, Allendale, MI 49401, USA}

\maketitle

\begin{abstract}
The \textsl{Fermi} LAT has detected numerous transient gamma-ray sources near the Galactic plane,
several of which have been shown to be located within our Galaxy.
We present an analysis of LAT Pass 8 data of seven previously reported, but still unidentified
transient gamma-ray sources located within $10^\circ$ of the Galactic plane.
We detect significant gamma-ray emission lasting several days for three of these sources:
Fermi J0035+6131, Fermi J0905-3527, and Fermi J0910-5041.
However, we were not able to detect the increase in gamma-ray emission that has previously been
reported in the other four cases.
We also review available multiwavelength data for the transients and discuss potential counterparts.
\end{abstract}

\section{INTRODUCTION}

Since the launch of the \textsl{Fermi} Gamma-ray Space Telescope, the Large Area Telescope (LAT)
has detected numerous transient sources of GeV gamma rays near the Galactic plane.
Some of these transients may represent previously unknown gamma-ray sources in our Galaxy.
The first \textsl{Fermi} LAT transient for which a Galactic origin could be demonstrated
was Fermi J2102+4542 \cite{atel2487}.
It was identified as a classical nova outburst in the symbiotic star V407 Cyg \cite{abdo2010}.
This detection established novae as a new class of gamma-ray sources
and is the first detection of GeV gamma rays from a symbiotic star.
Subsequently, the \textsl{Fermi} LAT has detected gamma-ray emission from several other novae:
Nova Sco 2012, Nova Mon 2012, Nova Del 2013, Nova Cen 2013, and Nova Sgr 2015.
However, several of the transients discovered near the Galactic plane were found to be
extragalactic sources serendipitously located behind the Galactic disk,
such as Fermi J0109+6134 which was identified as a blazar \cite{vandenbroucke2010}.
In order to determine whether a transient is Galactic or extragalactic,
it is necessary to identify and investigate counterparts at other wavelengths.
Of particular interest are X-ray and radio counterparts because very few are typically found
inside the \textsl{Fermi} LAT confidence region and their positions can be determined with much higher
accuracy than is possible with the gamma-ray data.

We investigated seven \textsl{Fermi} LAT transients located within $10^\circ$ of the Galactic plane
that have previously been reported in Astronomer's Telegrams but for which a Galactic or
extragalactic origin has not yet been established.
Here we present preliminary results of our analysis of LAT Pass 8 data of these transients
and our search for multiwavelength counterparts using archival data as well as new observations.
The transients considered here and some of their properties are shown in Table~1.

\begin{table}[ht]
\caption{Previously reported, but still unidentified transient \textsl{Fermi} LAT sources
located within $10^\circ$ of the Galactic plane.}
\label{transient_list}
\begin{tabular}{cccccc}
\hline
\bf{\textsl{Fermi} LAT Transient}  &  \multicolumn{2}{c}{\bf{Galactic Coordinates}}  &
                                      \bf{Duration}  &  \bf{Flux ($>$100 MeV)}  &  \bf{Reference}  \\
                                   &  \bf{Latitude}  &  \bf{Longitude}               &
                                                     &  \bf{(photons cm$^{-2}$ s$^{-1}$)}            &  \\
\hline
J0035+6131  &  $121.05^\circ$  &  $-1.29^\circ$  &  $\sim$2 days  &  $5.7\times10^{-7}$  &  ATel \#8554 \cite{atel8554}  \\
J0902-4624  &  $267.47^\circ$  &  $+0.07^\circ$  &  not reported  &  $1.2\times10^{-6}$  &  ATel \#3972 \cite{atel3972}  \\
J0905-3527  &  $259.59^\circ$  &  $+7.73^\circ$  &  $\sim$3 days  &  $1.5\times10^{-6}$  &  ATel \#1771 \cite{atel1771}  \\
J0910-5041  &  $271.62^\circ$  &  $-1.80^\circ$  &  $\sim$2 days  &  $1.4\times10^{-6}$  &  ATel \#1788 \cite{atel1788}  \\
J1057-6027  &  $289.30^\circ$  &  $-0.64^\circ$  &  not reported  &  $2.4\times10^{-6}$  &  ATel \#2081 \cite{atel2081}  \\
J1643-4558  &  $339.05^\circ$  &   $0.00^\circ$  &  $\sim$1 hour  &  $4.3\times10^{-6}$  &  ATel \#4285 \cite{atel4285}  \\
J1725-3726  &  $350.55^\circ$  &  $-1.02^\circ$  &  $\sim$5 min   &  $1.3\times10^{-5}$  &  ATel \#5625 \cite{atel5625}  \\
\hline
\end{tabular}
\end{table}

\newpage

\section{FERMI J0035+6131}

\begin{figure}[ht]
\includegraphics[height=4.6cm]{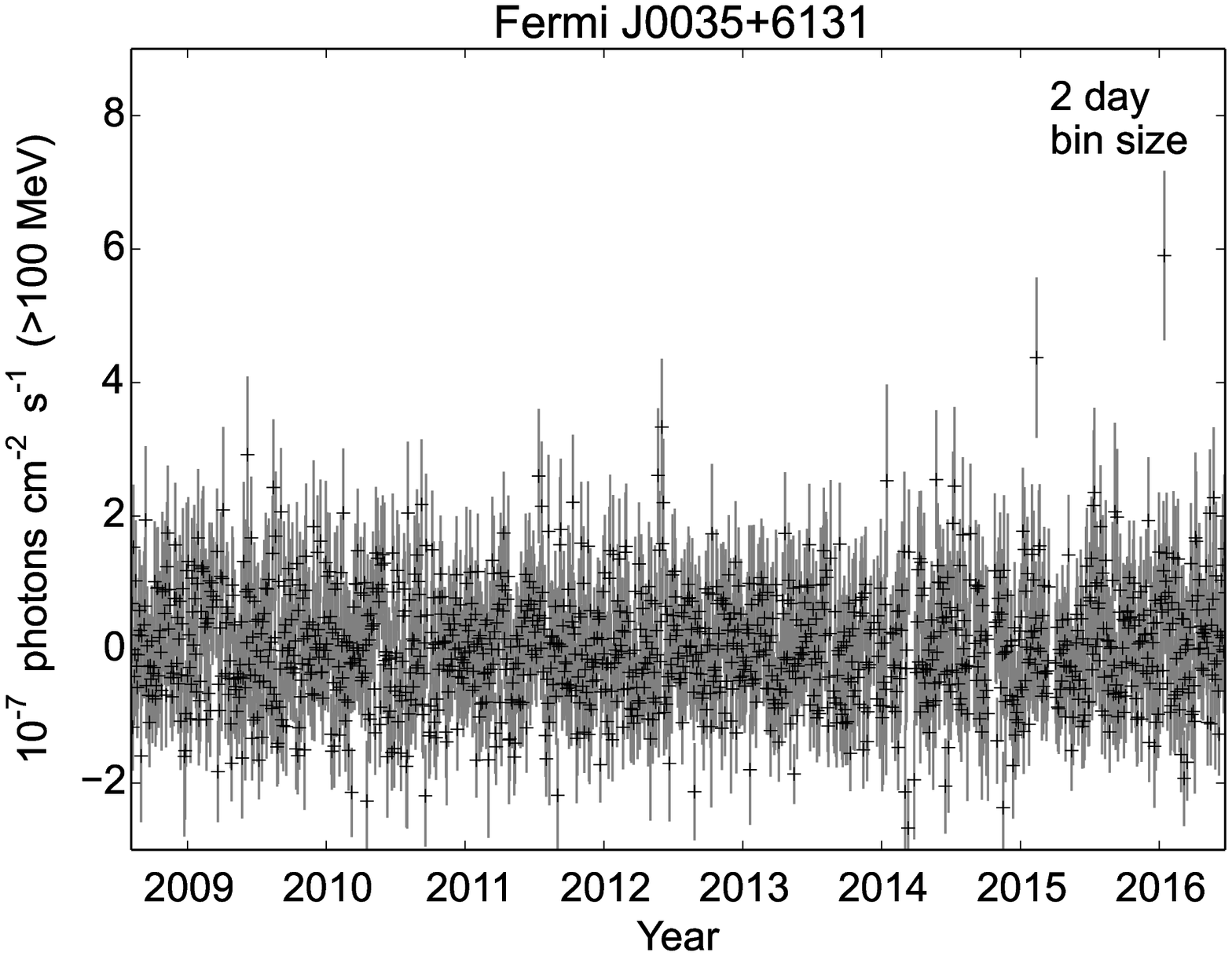}
\hspace{0.5cm}
\includegraphics[height=4.6cm]{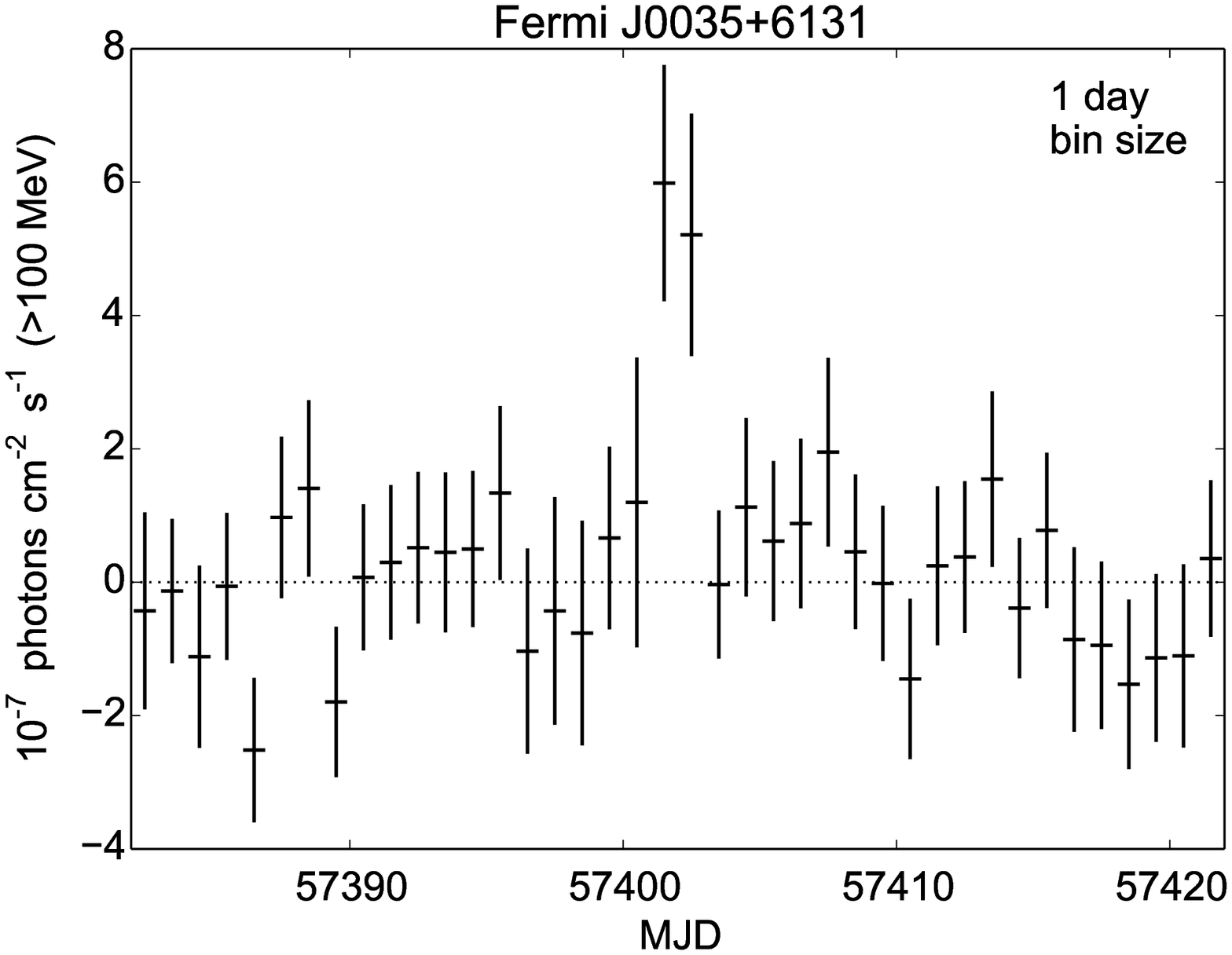}
\caption{
\textsl{Fermi} LAT light curve ($>$100 MeV) of Fermi J0035+6131 obtained using a circular aperture
with a $1^\circ$ radius.
\textsl{(Left)}~8~year light curve with a 2 day bin size.
\textsl{(Right)} Daily light curve around the time of the outburst.
}
\label{j0035a}
\end{figure}

\begin{figure}[ht]
\begin{tabular}{p{0.8cm} p{7.5cm} p{7cm}}
&
\vspace{-5pt} \includegraphics[height=6.4cm]{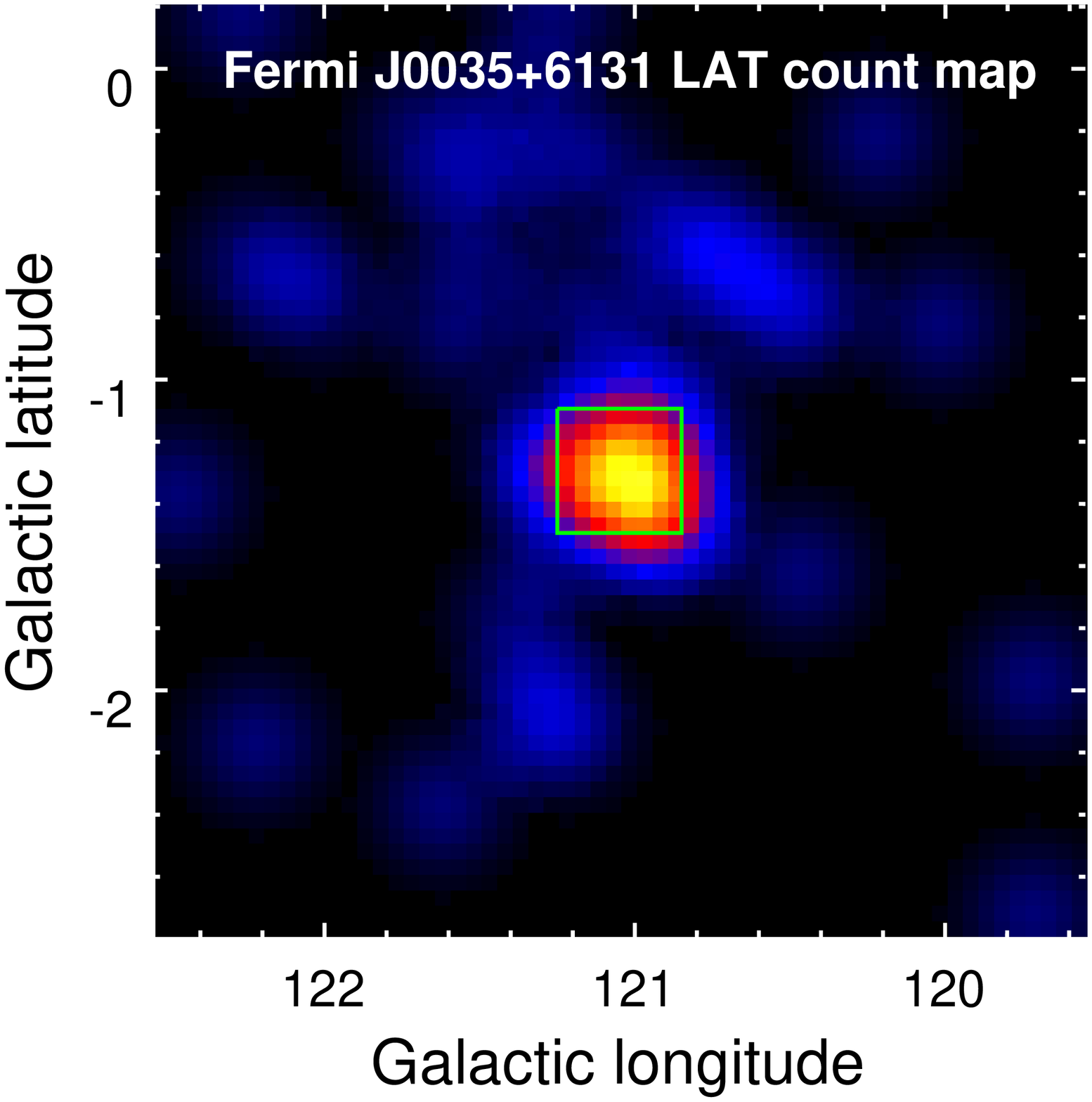} &
\vspace{-2pt} \includegraphics[height=5.4cm]{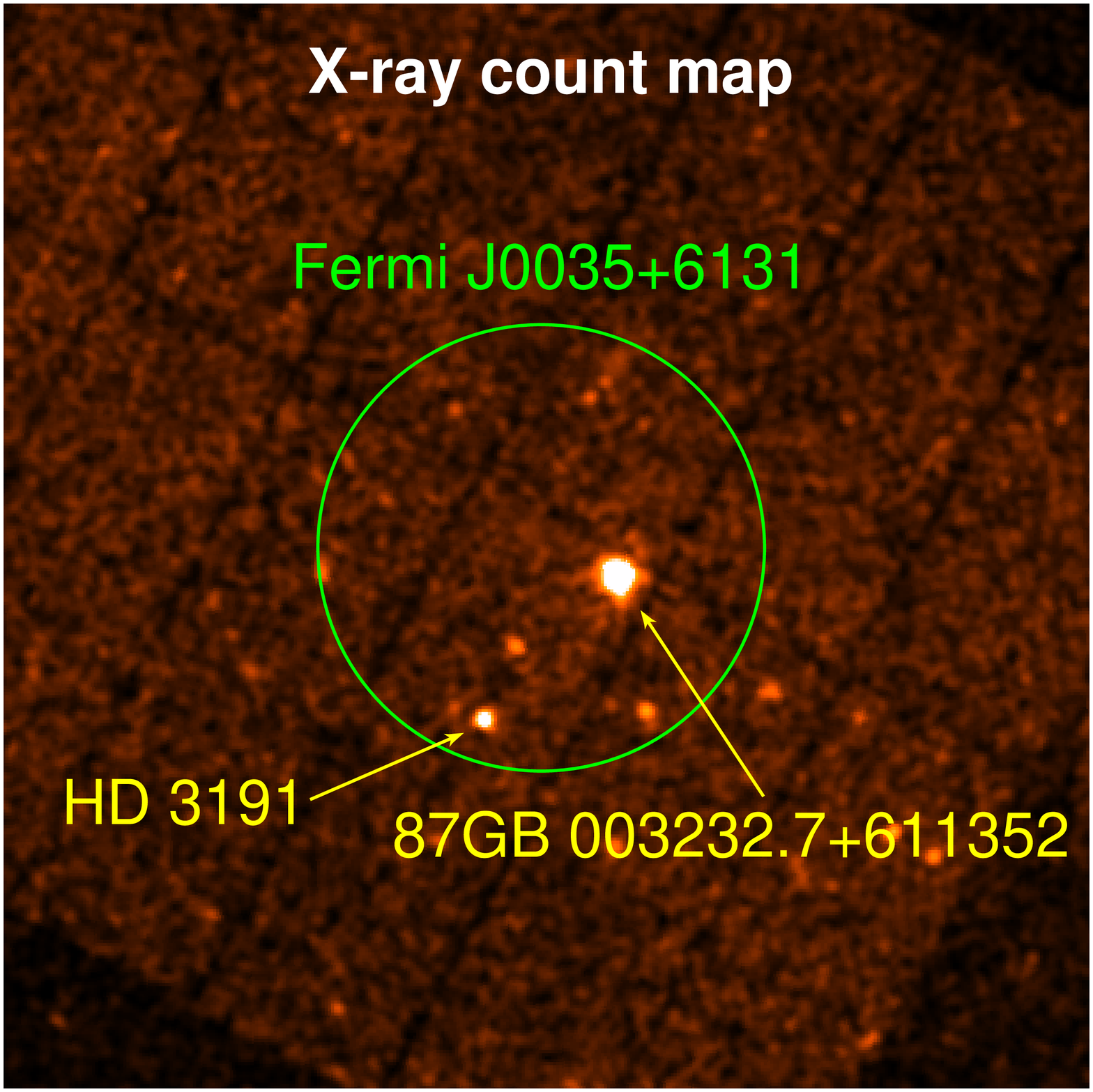}
\end{tabular}
\caption{
\textsl{(Left)} \textsl{Fermi} LAT count map ($>$500 MeV) of Fermi J0035+6131 during the 2 day outburst.
\textsl{(Right)} \textsl{XMM-Newton} X-ray count map of a $0.4^\circ\times0.4^\circ$ region around
Fermi J0035+6131 (shown as a square on the LAT count map).
}
\label{j0035b}
\end{figure}

The \textsl{Fermi} LAT detected the transient source Fermi J0035+6131 on January 14, 2016
(MJD 57401) \cite{atel8554} with the gamma-ray outburst lasting about two days (Figure~\ref{j0035a}).
No known gamma-ray source at this location is listed in the LAT 4-year point source catalog \cite{acero2015}.
The transient was observed 10 days later with \textsl{XMM-Newton} (Figure~\ref{j0035b}) which revealed
two potential \mbox{X-ray} counterparts inside the \textsl{Fermi} LAT confidence region \cite{atel8783}.
One of these X-ray sources is associated with the unidentified radio source 87GB 003232.7+611352
(NVSS J003524+613030).
This source currently does not have a known optical counterpart at visual wavelengths,
although a detection in the near infrared has been reported \cite{atel8706}.
The radio emission and the low optical brightness suggest that the source may be a blazar
serendipitously located behind the Galactic disk.
However, additional optical observations are needed to determine, based on redshift
measurements, whether the source is Galactic or extragalactic.
The other X-ray source inside the \textsl{Fermi} LAT confidence region is associated with
the B1~IV star HD~3191.
The source exhibits a hard \mbox{X-ray} spectrum which suggests that it may be a
High-mass X-ray Binary (HMXB).
This interpretation is supported by the optical spectrum which shows a large stellar rotation velocity
and a change in the radial velocity compared to historical data \cite{atel8789}.
Further radial velocity measurements are needed to search for any orbital motion of the companion star
around the compact object.
HD~3191 is an interesting potential counterpart of the gamma-ray transient
as \textsl{Fermi} has detected variable GeV emission from several X-ray binaries, such as
the microquasar Cygnus X-3 \cite{abdo2009a} and the
High-mass X-ray Binaries LSI~+61$^\circ$303 and LS~5039 \cite{abdo2009b}.

\newpage

\section{FERMI J0905-3527}

\begin{figure}[ht]
\includegraphics[height=4.6cm]{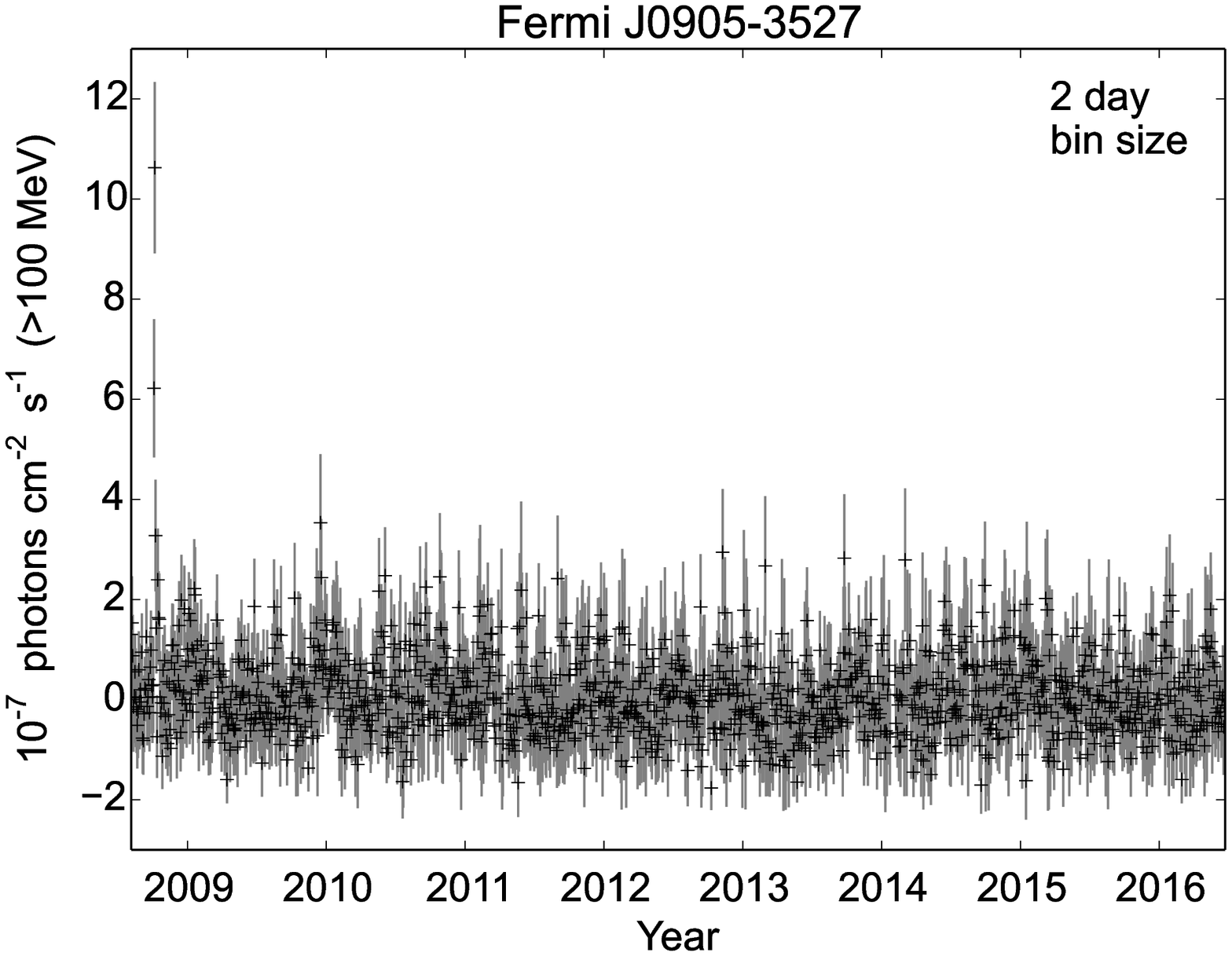}
\hspace{0.5cm}
\includegraphics[height=4.6cm]{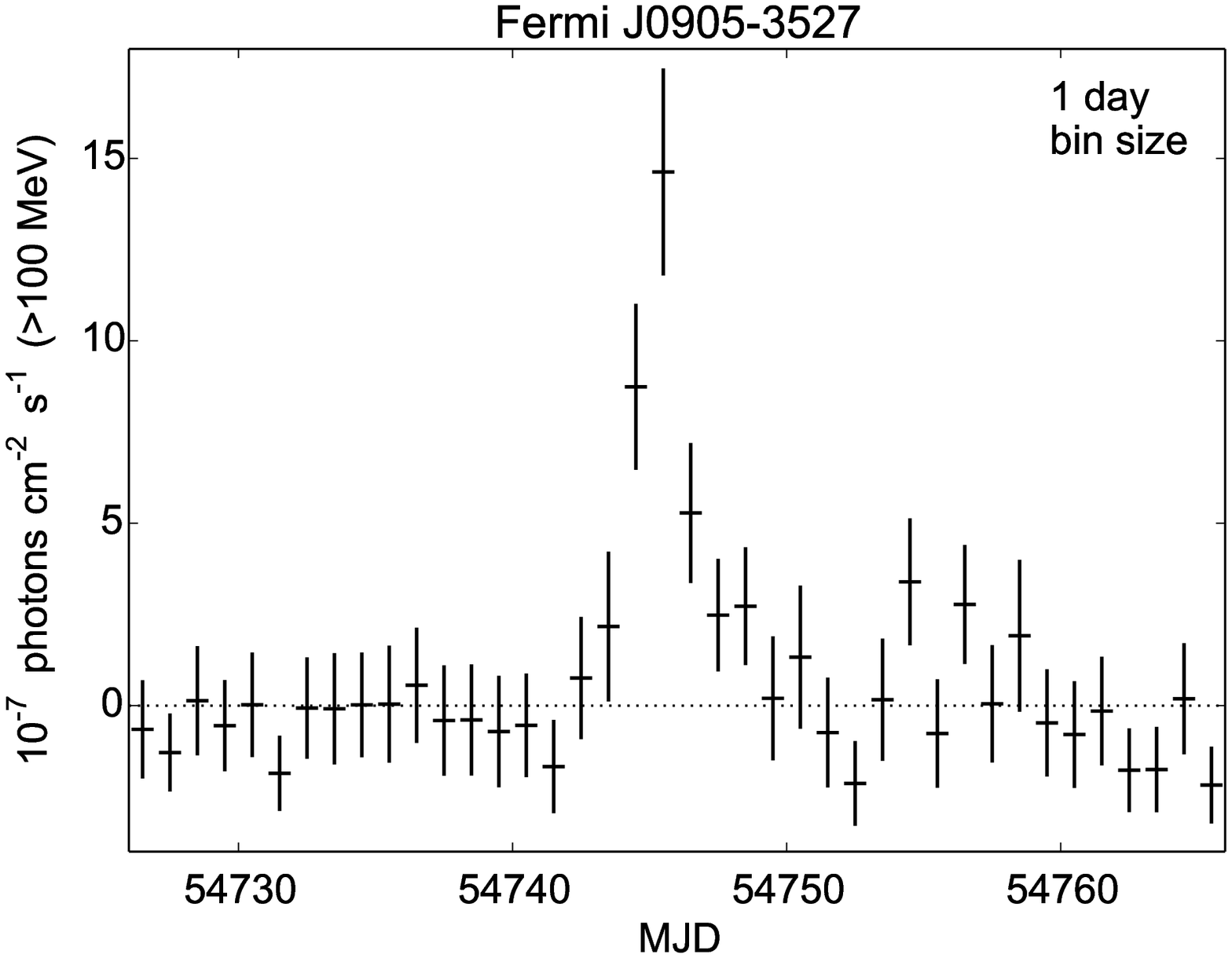}
\caption{
\textsl{Fermi} LAT light curve ($>$100 MeV) of Fermi J0905-3527 obtained using a circular aperture
with a $1^\circ$ radius.
\textsl{(Left)}~8~year light curve with a 2 day bin size.
\textsl{(Right)} Daily light curve around the time of the outburst.
}
\label{j0905a}
\end{figure}

\begin{figure}[ht]
\begin{tabular}{p{0.8cm} p{7.5cm} p{7cm}}
&
\vspace{-5pt} \includegraphics[height=6.4cm]{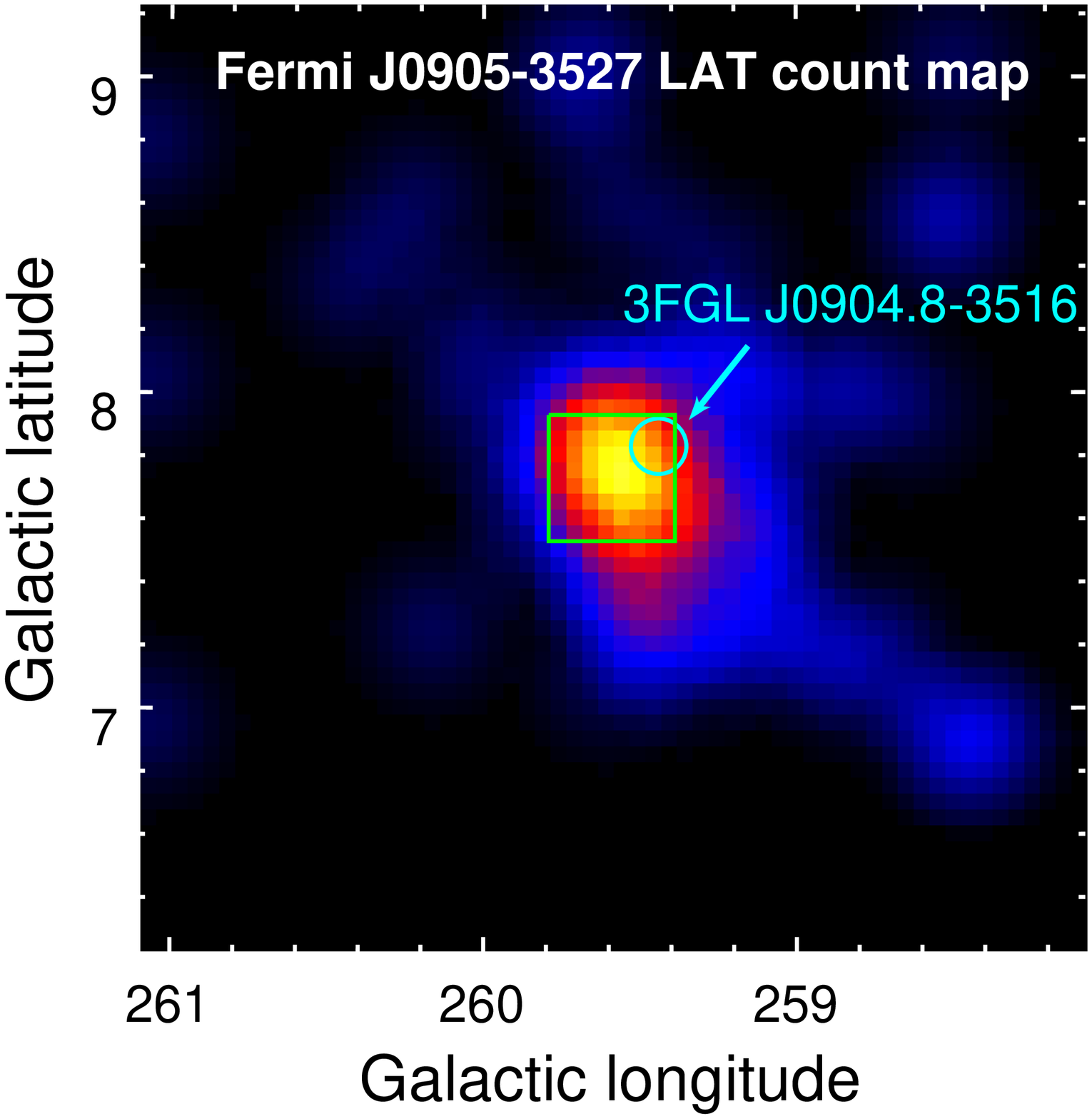} &
\vspace{-2pt} \includegraphics[height=5.4cm]{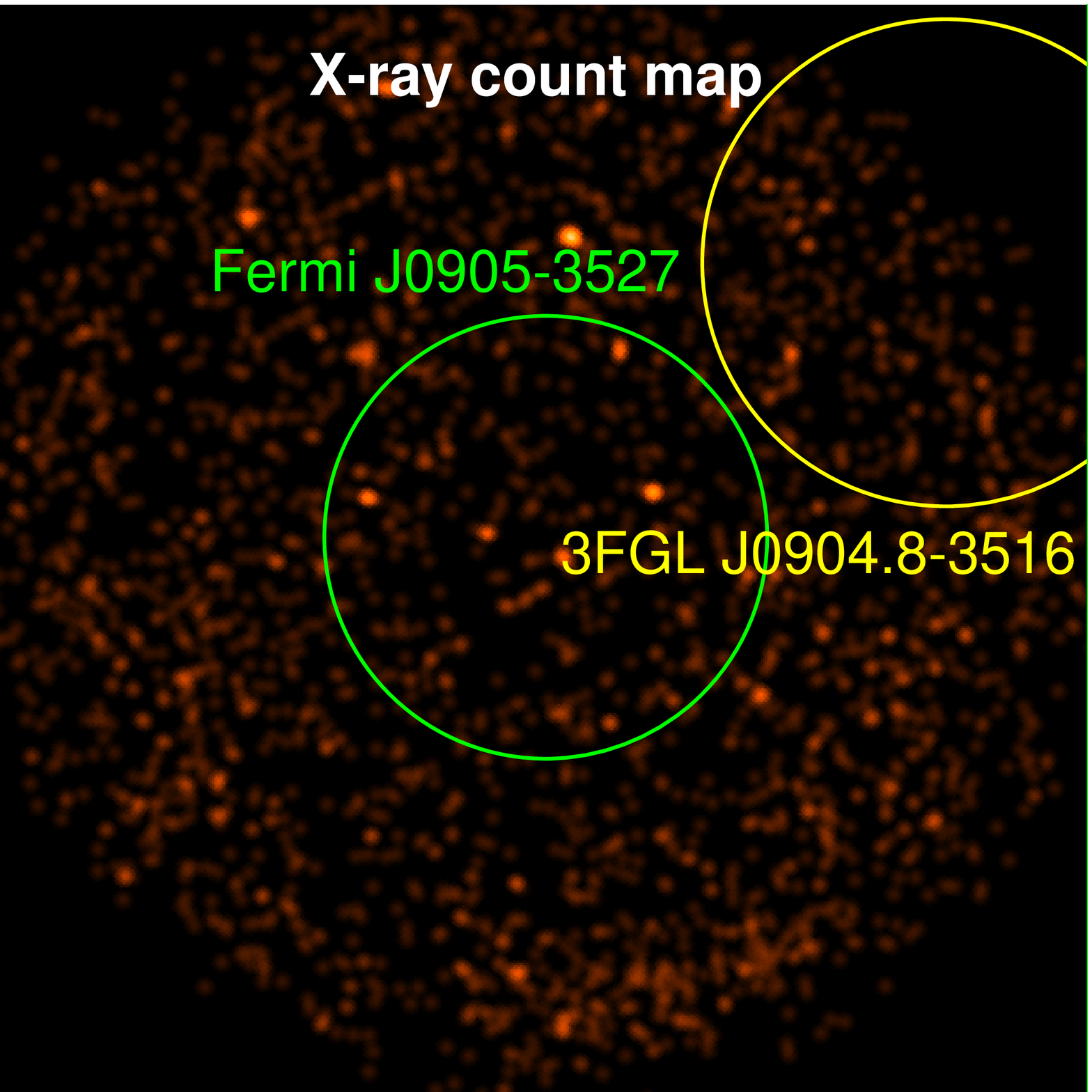}
\end{tabular}
\caption{
\textsl{(Left)} \textsl{Fermi} LAT count map ($>$500 MeV) of Fermi J0905-3527 during the 3 day outburst.
\textsl{(Right)} \textsl{Swift} X-ray count map of a $0.4^\circ\times0.4^\circ$ region around
Fermi J0905-3527 (shown as a square on the LAT count map).
}
\label{j0905b}
\end{figure}

The transient gamma-ray source Fermi J0905-3527 was detected on October 6, 2008 (MJD 54745)
\cite{atel1771} and lasted about three days (Figure~\ref{j0905a}).
It was initially thought to be associated with the nearby EGRET source 3EG J0903-3531.
The LAT 4-year point source catalog also lists the nearby source 3FGL J0904.8-3516
\cite{acero2015} which appears to be associated with the radio source NVSS J090442-351423.
However, our preliminary analysis of the \textsl{Fermi} LAT data suggests that the position
of Fermi J0905-3527 is not consistent with 3FGL J0904.8-3516 or the radio source NVSS J090442-351423,
indicating that Fermi J0905-3527 may be a distinct and as yet unidentified gamma-ray source,
although further analysis is needed to rule out any systematic errors in the source position.
The region around Fermi J0905-3527 was also observed with \textsl{Swift} (Figure~\ref{j0905b}).
While several marginally detected X-ray source are found inside the LAT confidence region,
none of them can be clearly identified as a counterpart or associated with sources
at other wavelengths.
The only potential radio counterpart consistent with the LAT confidence region of the transient
is the faint, unidentified radio source NVSS J090458-353145.

\newpage

\section{FERMI J0910-5041}

\begin{figure}[ht]
\includegraphics[height=4.6cm]{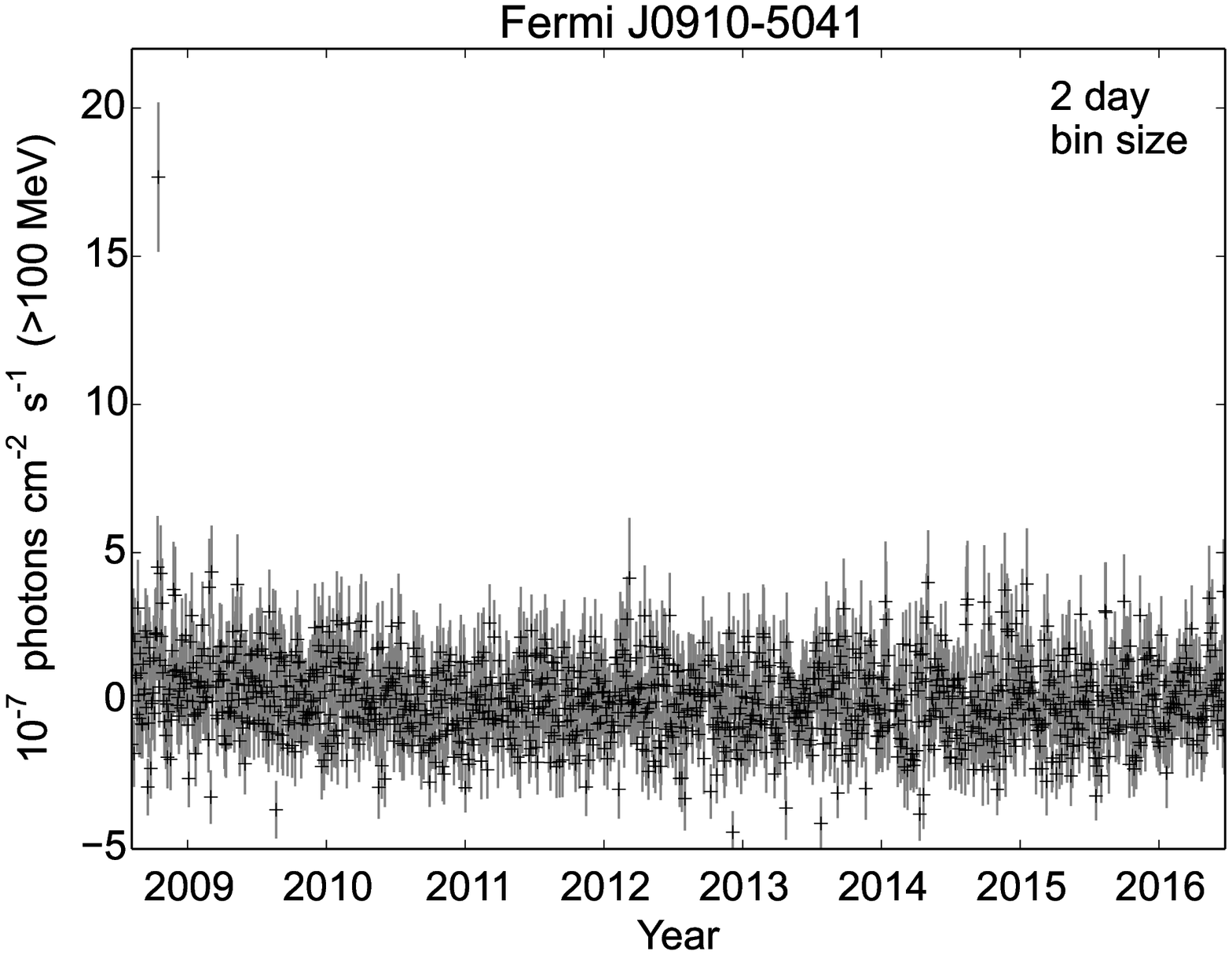}
\hspace{0.5cm}
\includegraphics[height=4.6cm]{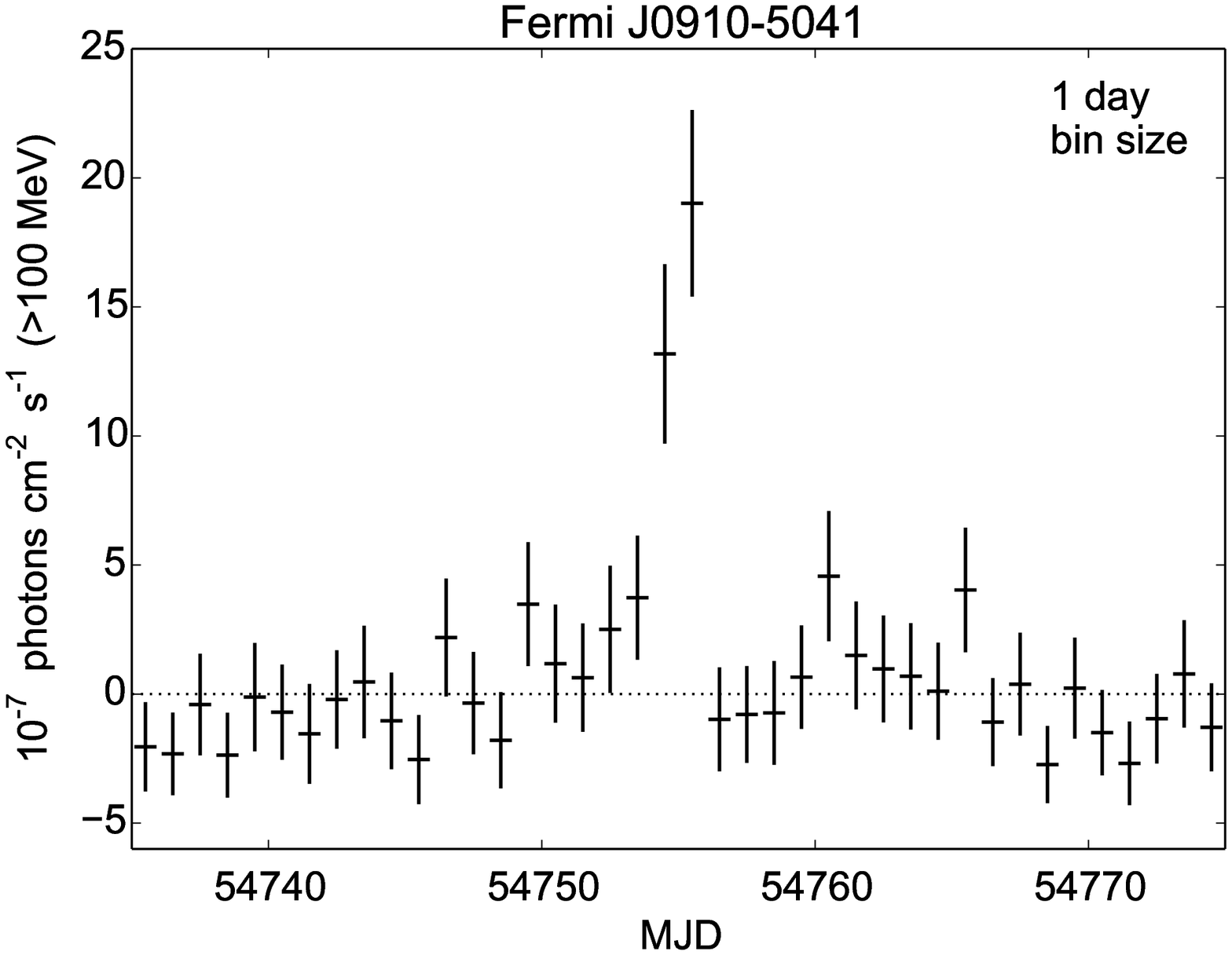}
\caption{
\textsl{Fermi} LAT light curve ($>$100 MeV) of Fermi J0910-5041 obtained using a circular aperture
with a $1^\circ$ radius.
\textsl{(Left)}~8~year light curve with a 2 day bin size.
\textsl{(Right)} Daily light curve around the time of the outburst.
}
\label{j0910a}
\end{figure}

\begin{figure}[ht]
\begin{tabular}{p{0.8cm} p{7.5cm} p{7cm}}
&
\vspace{-5pt} \includegraphics[height=6.4cm]{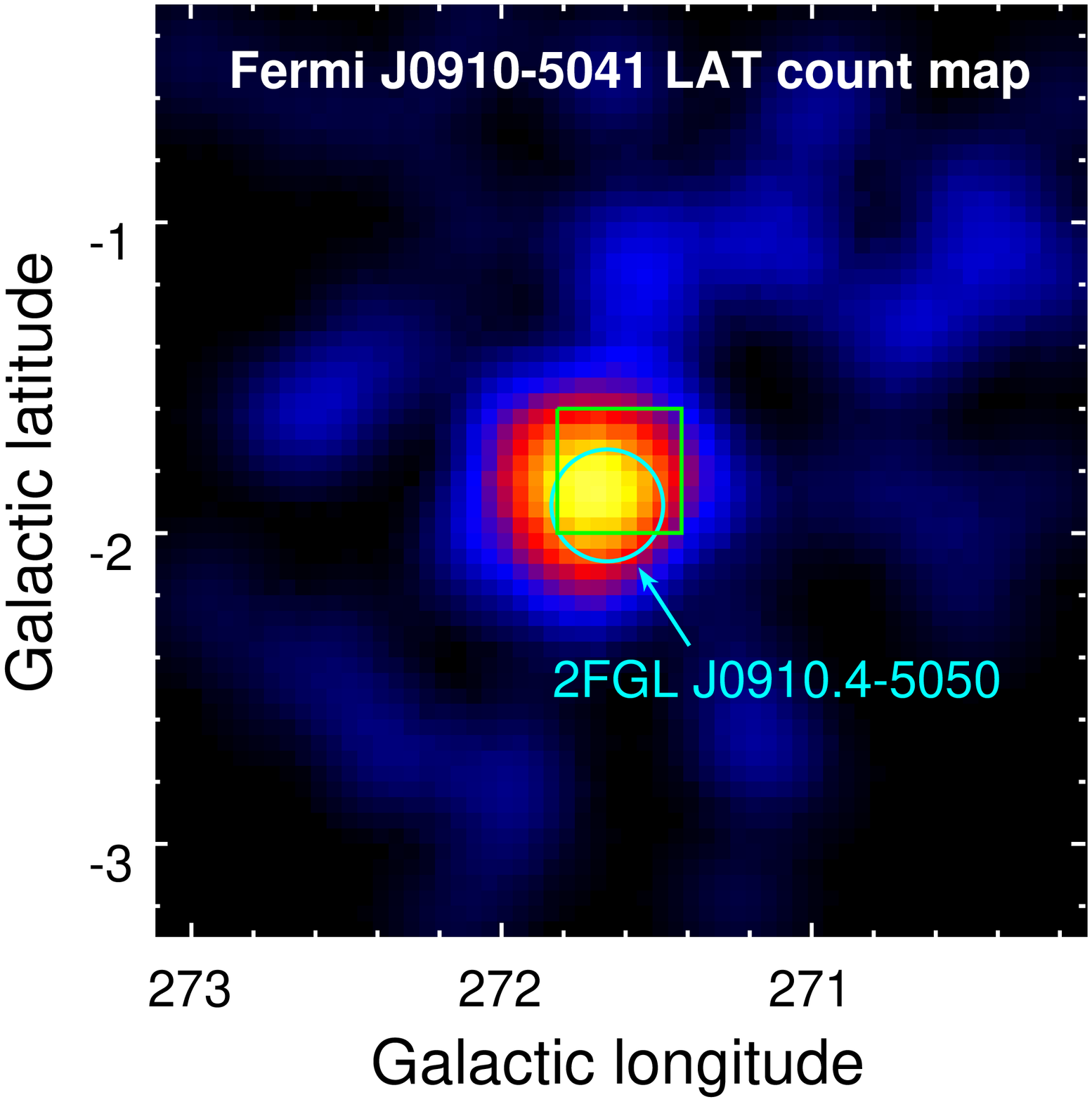} &
\vspace{-2pt} \includegraphics[height=5.4cm]{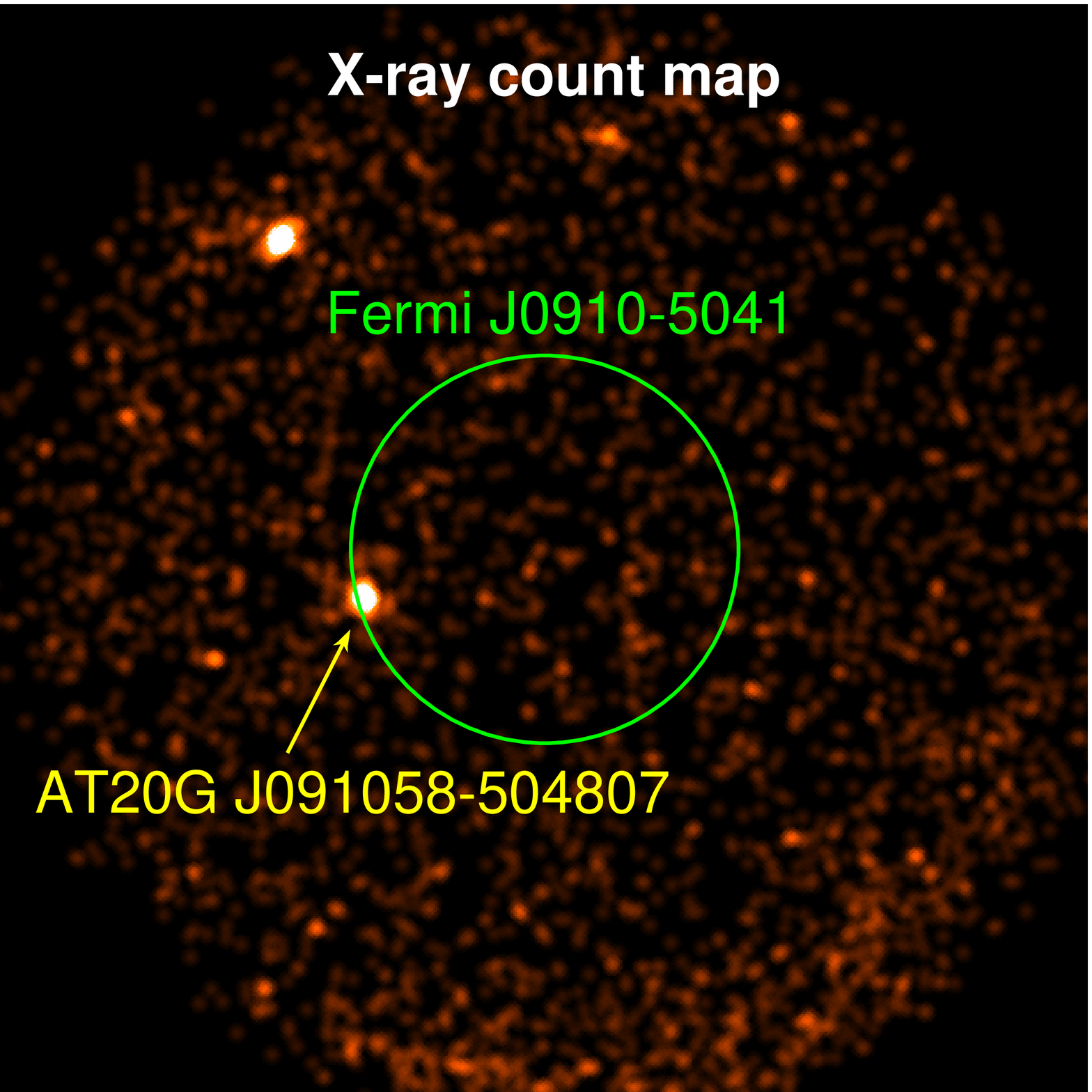}
\end{tabular}
\caption{
\textsl{(Left)} \textsl{Fermi} LAT count map ($>$500 MeV) of Fermi J0910-5041 during the 2 day outburst.
\textsl{(Right)} \textsl{Swift} X-ray count map of a $0.4^\circ\times0.4^\circ$ region around
Fermi J0910-5041 (shown as a square on the LAT count map).
}
\label{j0910b}
\end{figure}

Fermi J0910-5041 was detected on October 15, 2008 (MJD 54754) \cite{atel1788}
with the gamma-ray outburst lasting about two days (Figure~\ref{j0910a}).
The transient is positionally coincident with the gamma-ray source 2FGL J0910.4-5050 listed
in the LAT 2-year point source catalog \cite{nolan2012}.
However, this source is not listed in the LAT 4-year point source catalog \cite{acero2015}
and may have been a false detection caused by the strong, diffuse Galactic gamma-ray emission
present in this region.
Our preliminary analysis of 8 years of LAT Pass 8 data also does not show a gamma-ray
point source at this position outside the 2-day time period of the Fermi J0910-5041 outburst.
\textsl{Swift} observations performed a few days after the gamma-ray discovery (Figure~\ref{j0910b})
revealed one likely X-ray counterpart inside the LAT confidence region \cite{atel1822}.
The X-ray source appears to be associated with the radio source AT20G J091058-504807 \cite{atel1843}.
No optical counterpart is found at the position of the X-ray or radio source.
If Fermi J0910-5041 is associated with AT20G J091058-504807, the strong X-ray and radio emission
and the lack of optical emission suggest that the gamma-ray source is an AGN
serendipitously located behind the Galactic disk.

\newpage

\section{FERMI J0902-4624, J1057-6027, J1643-4558, AND J1725-3726}

\begin{figure}[ht]
\begin{tabular}{ccc}
\includegraphics[height=4.6cm]{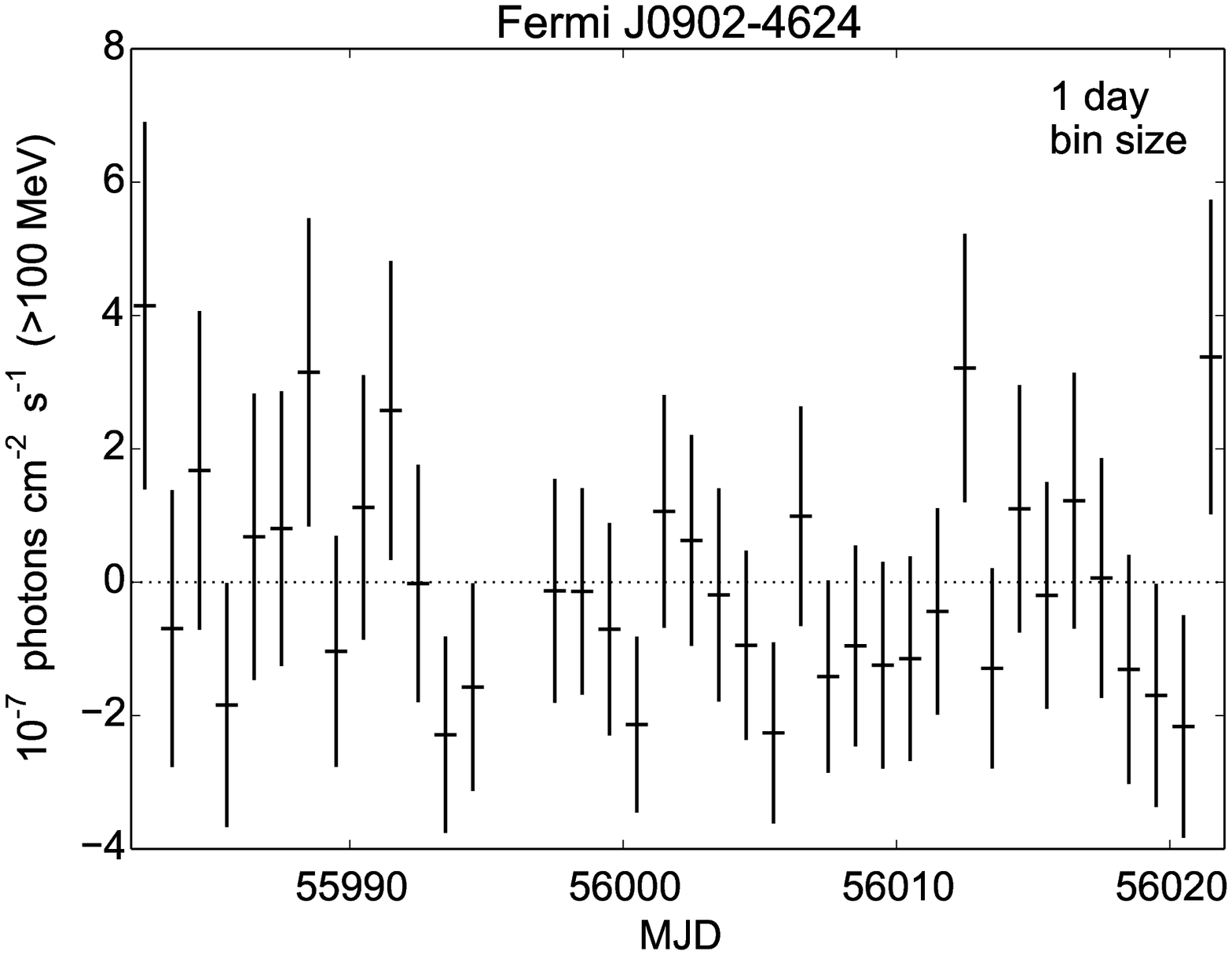} & &
\includegraphics[height=4.6cm]{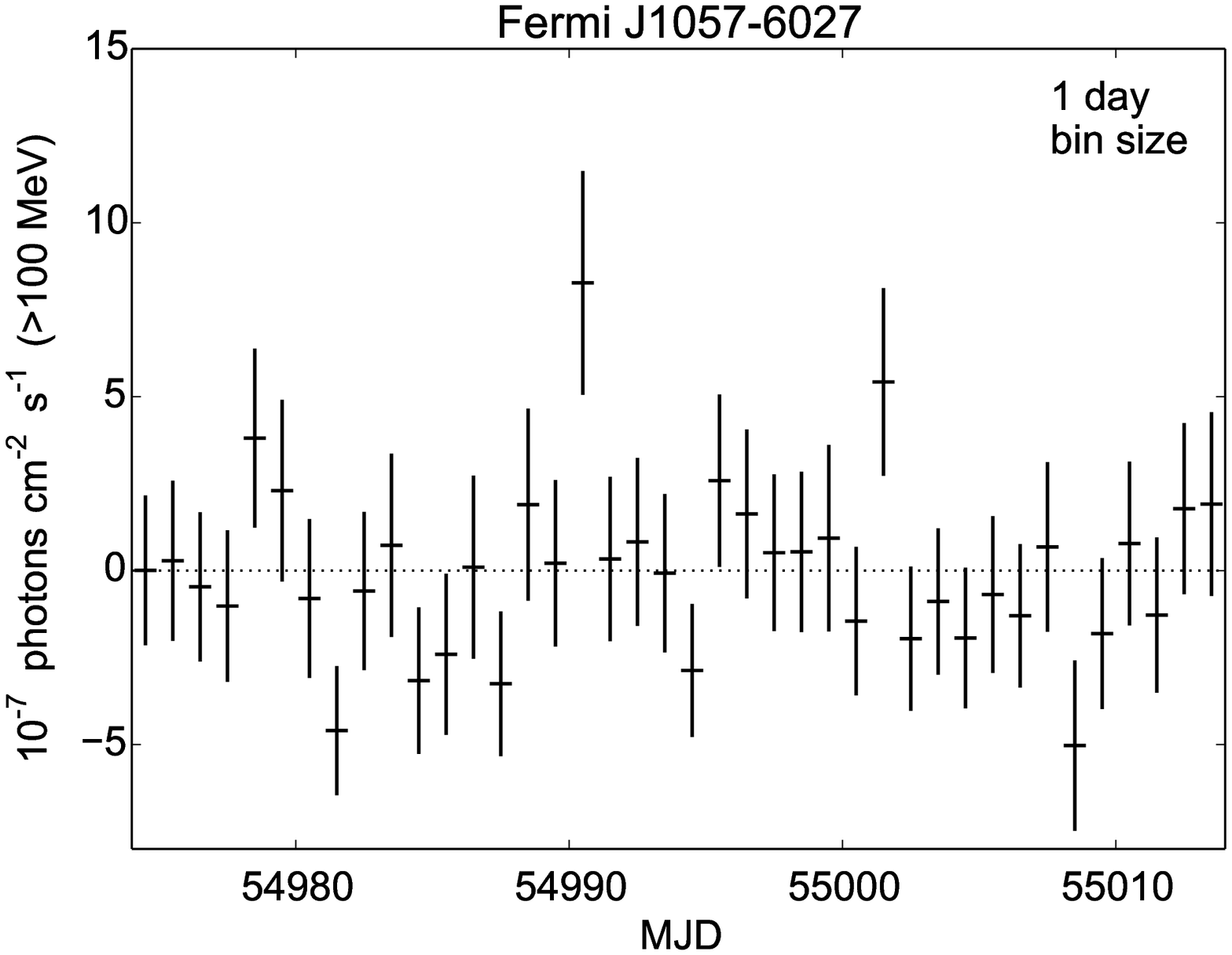} \\
\includegraphics[height=4.6cm]{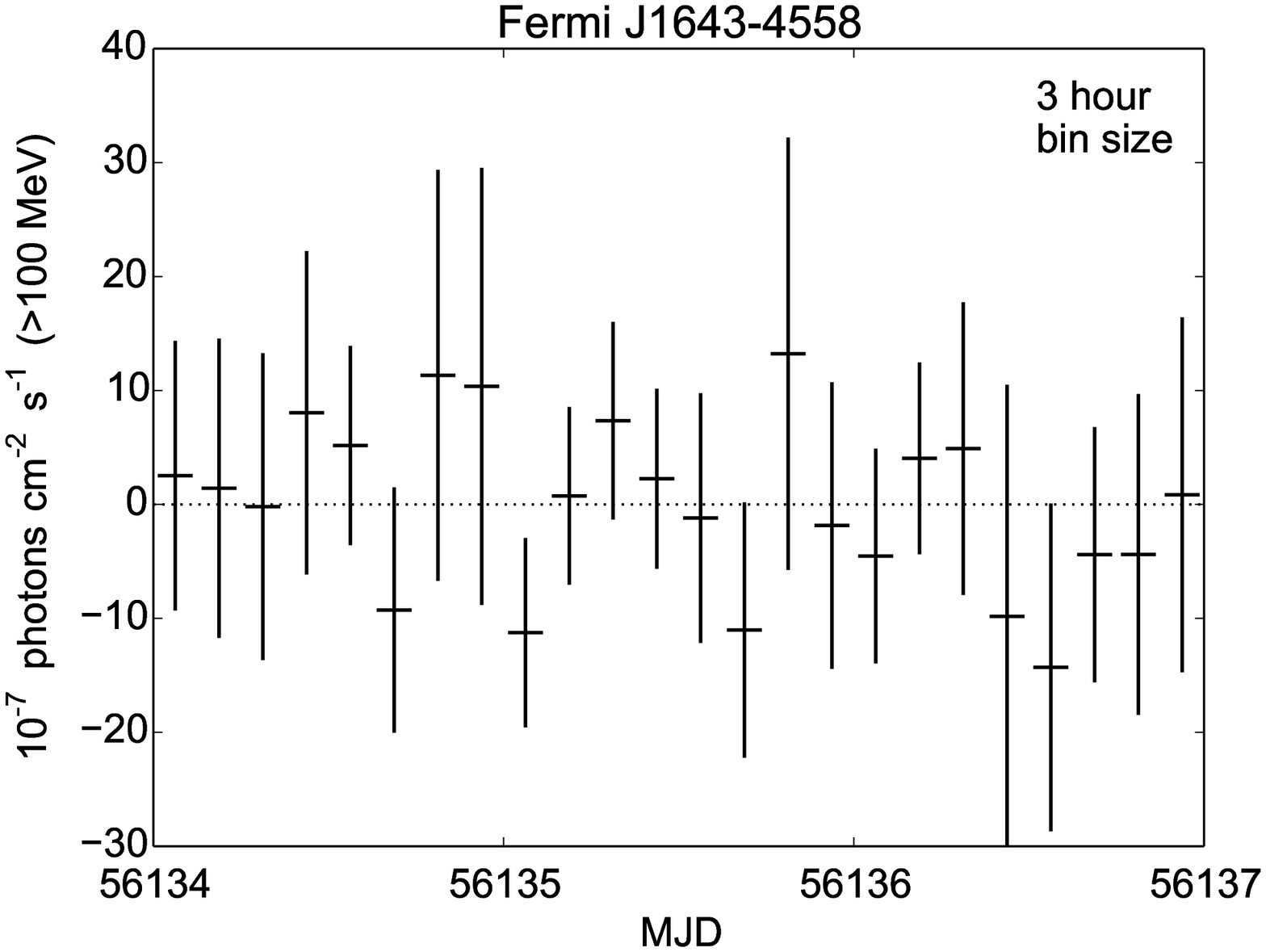} & &
\includegraphics[height=4.6cm]{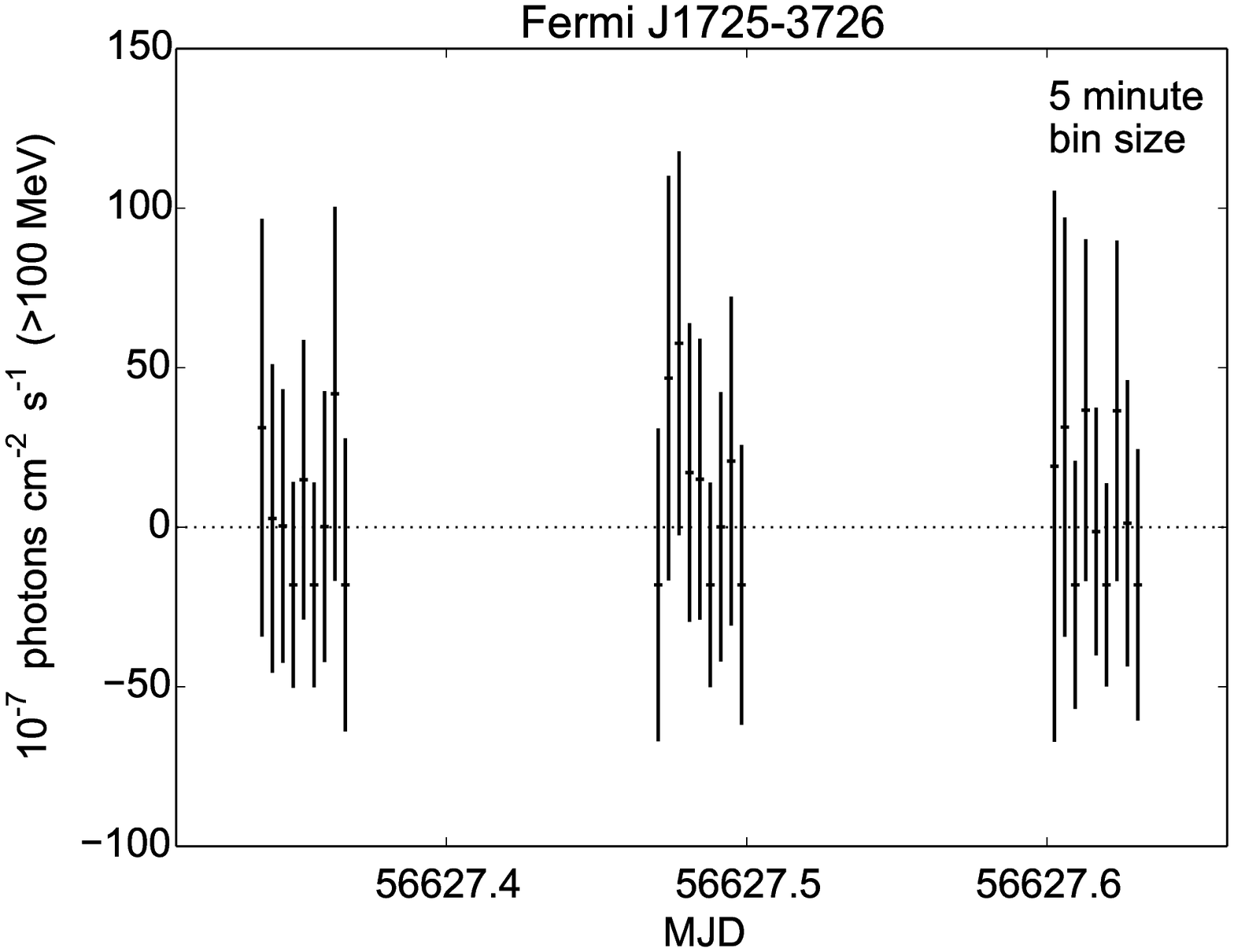}
\end{tabular}
\caption{\textsl{Fermi} LAT light curves ($>$100 MeV) for the time periods and positions
originally reported for the transients Fermi J0902-4624, Fermi J1057-6027, Fermi J1643-4558,
and Fermi J1725-3726.}
\label{lc_other}
\end{figure}

We analyzed the LAT Pass 8 data for the time periods and positions originally reported
for the other four transients, but we were not able to detect any significant gamma-ray excess
(Figure \ref{lc_other}).
It is possible that the improvements in the event reconstruction implemented for the Pass 8
data release resulted in the rejection of the photon events that were originally reported
for these transients.
For our analysis of the LAT data, we included front and back events classified as ''source''
(\textsl{evclass}=128) with a zenith angle $\le$90$^\circ$ as is recommended for a standard analysis.
We note that these transient sources have also not been detect by the
\textsl{Fermi} All-sky Variability Analysis (FAVA) \cite{ackermann2013}.

Fermi J0902-4624 was detected on March 15, 2012 (MJD 56001) \cite{atel3972}.
No gamma-ray source consistent with this location is listed in the LAT 4-year
point source catalog \cite{acero2015}.
Follow-up observations with \textsl{Swift} revealed four X-ray sources inside
the LAT confidence region \cite{atel3973}, three of which were found to be variable \cite{atel3992}.
However, no clear identification of an X-ray counterpart of the gamma-ray transient has been made.

Fermi J1057-6027 was detected on June 11, 2009 (MJD 54993) \cite{atel2081}.
The closest gamma-ray source listed in the LAT 4-year point source catalog is
3FGL J1055.8-6025  \cite{acero2015} which is associated with the pulsar PSR J1055-6028.
However, this source is $0.17^\circ$ offset from the best-fit position of Fermi J1057-6027
and outside the $0.068^\circ$ radius LAT confidence region originally reported.
Given the non-detection of the transient event in the LAT Pass 8 data, the gamma-ray
emission originally reported is likely not associated with PSR J1055-6028.
Follow-up observations with \textsl{Swift} did not reveal any X-ray sources inside
the LAT confidence region of Fermi J1057-6027 \cite{atel2083}.

Fermi J1643-4558 was detected on July 27, 2012 (MJD 56135.31) during one orbit
of the satellite, which implies a duration between 0.5 and 2 hours \cite{atel4285}.
The closest known gamma-ray source listed in the LAT 4-year point source catalog
is 3FGL J1641.1-4619c \cite{acero2015} with an offset of $0.5^\circ$,
which is near the edge, but outside the 95\% LAT confidence region of the transient's position.
Also notable is the gamma-ray pulsar PSR J1648-4611 which is offset by $0.9^\circ$
from the reported transient position.

Fermi J1725-3726 was detected on December 1, 2013 (MJD 56627.474)
with a duration of only 5 minutes \cite{atel5625}.
The LAT 4-year point source catalog does not list any gamma-ray sources consistent
with the location of the transient \cite{acero2015}.
A notable X-ray source coincident with the transient's position is the X-ray burster XTE J1723-376
located near the edge of the 95\% LAT confidence region.

\newpage

\section{SUMMARY}

We have analyzed \textsl{Fermi} LAT Pass 8 data of seven previously reported, but still unidentified
transient gamma-ray sources located within $10^\circ$ of the Galactic plane
and reviewed available multiwavelength data to identify potential counterparts.
For four of the transients, we were unable to detect the increase in gamma-ray emission that has
previously been reported.
For the other three transients, we detect significant gamma-ray emission lasting 2--3 days.
While the multiwavelength properties of Fermi J0910-5041 suggest that it is an AGN
serendipitously located behind the Galactic disk,
further observations of the other two transients are needed to determine whether
they are Galactic or extragalactic.
For Fermi J0035+6131, we found two potential X-ray counterparts, one of which may be a
High-mass X-ray Binary in our Galaxy, while the other is likely a blazar.
For Fermi J0905-3527, our preliminary analysis indicates that it is not associated
with the nearby gamma-ray and radio sources that were previously suggested as counterparts.
Further analysis of the \textsl{Fermi} LAT data is needed to rule out systematic errors
in the position of Fermi J0905-3527, and additional multiwavelength observations are necessary to identify
potential counterparts consistent with the LAT confidence region.

\section{ACKNOWLEDGMENTS}
D.~Pandel acknowledges support from NASA Guest Investigator Grant 15AJ88G.
This work is based on observations obtained with \textit{XMM-Newton}, an ESA science mission
with instruments and contributions directly funded by ESA member states and NASA.

\nocite{*}
\bibliographystyle{aipnum-cp}
\bibliography{fermi_transients}

\end{document}